\input xy
\xyoption{all}

\documentclass[11pt]{article}
\input epsf.tex
\usepackage{pstricks}
\usepackage{psfrag}
\usepackage{graphicx}
\usepackage[small,loose]{subfigure}
\usepackage[latin2]{inputenc}
\usepackage{epic}
\usepackage{latexsym}
\usepackage{mathrsfs}
\usepackage{bbm} % BlackBoeard letters
\usepackage{cancel} 
\usepackage{amssymb,amsmath}
\usepackage{fleqn} 
\def\harr#1#2{\smash{\mathop{\hbox to .3in{\rightarrowfill}}
 \limits^{\scriptstyle#1}_{\scriptstyle#2}}}

\def\yzero{\smash{\hbox{$y\kern-4pt\raise1pt\hbox{${}^\circ$}$}}}

\def\sin2{\frac{1}{\sqrt2}}

\def\wt{\widetilde}
\def\be{\begin{equation}}
\def\ee{\end{equation}}
\def\beqa{\begin{eqnarray}}
\def\eeqa{\end{eqnarray}}
\def\bl{\begin{eqnarray}}
\def\el{\end{eqnarray}}

\def\Dsl{\,\raise.15ex\hbox{/}\mkern-13.5mu D} %can be subscripted

\def\bfr{\begin{flushright}}
\def\efr{\end{flushright}}
\def\wt{\widetilde}

\def\H{\not{\!\!H}}

\def\cov{\widetilde\nabla}

\def\*{\ast_{(6+1)}}

\newcommand{\nn}{\nonumber}

\topmargin
-.5cm
\textwidth
15.5cm
\textheight
23cm
\oddsidemargin
0.7cm
\evensidemargin
1.2cm

\begin{document}
%----------------------------------------------------------------------%
%  numbering equations with section number
%----------------------------------------------------------------------%
\makeatletter
\@addtoreset{equation}{section}
\makeatother
\renewcommand{\theequation}{\thesection.\arabic{equation}}

%-------------------------------------------------------------------------------------
%                               Title page
%_____________________________________________________________________________________

\rightline{UG-DCI-GFM/10-19}

\vspace{.5cm}
\begin{center}
\Large{\bf Near-Horizon geometry from flux compactification}\\

\vspace{1cm}

\large
Oscar Loaiza-Brito$^{}$\footnote{e-mail :{\tt oloaiza@fisica.ugto.mx}} and Liliana Vazquez-Mercado$^{}$\footnote{e-mail :{\tt livazquezm@fisica.ugto.mx}}\\[4mm]

%{\small \em Centro de Investigaci\'on y de Estudios Avanzados del I.P.N., Unidad Monterrey}\\
%{\small\em Autopista Monterrey-Aeropuerto Km 10, 66600 Apodaca, Nuevo Le\'on, M\'exico,}\\
%{\small and}\\
{\small\em $^a$ Departamento de F\'isica, DCI,  Campus Le\'on,}\\
{\small\em Universidad de Guanajuato, C.P. 37150, Guanuajuato, M\'exico.}\\[4mm]
\vspace*{2cm}
\small{\bf Abstract} \\
\end{center}

\begin{center} 
\begin{minipage}[h]{14.0cm} { We study the conditions an arbitrary flux configuration  must fulfill in order  to construct a 4d space-time of the type $AdS_2\times S^2$  from a type IIB supergravity flux compactification in which NS-NS fluxes are included. We present a solution consisting on a compactification in the presence of 3-form NS-NS and RR fluxes. The internal manifold is a $SU(3)$ structure six-dimensional manifold, with null curvature and with torsion.  By preserving two supersymmetries in the four-dimensional low energy theory, we find a  way to obtain the $AdS_2\times S^2$ geometry as a near-horizon solution by compactification in non-Calabi-Yau manifolds.}

\end{minipage} 
\end{center}

\bigskip

\bigskip
 
\bigskip

\newpage

%--------------------------------------------------------------------------------------
%                       Paper begins
%-------------------------------------------------------------------------------------

\section{Introduction}
Solutions of $AdS_2\times S^2$ as near-horizon geometries of extremal black holes have been studied in the past years in the context of ${\cal N}=2$, $d=4$ ungauged supergravity   by performing Calabi-Yau (CY) compactifications in type II superstring theories  (see \cite{Balasubramanian:1998ee} and references therein).  A typical example considers type IIB superstring theory compactified on a CY threefold on which D3-branes wrap internal supersymmetric three-cycles.  In the low energy four-dimensional (4d) effective theory, this is interpreted as an extremal supersymmetric black-hole.  The $AdS_2\times S^2$ geometry emerges once the corresponding near-horizon limit is taken. This is equivalent to consider a compactification of type IIB superstring to a 2-dimensional space-time in  a CY threefold $X_6$  times $S^2$ threaded with a 5-form RR flux of the form $F_5=\omega_2\wedge F_3$, where $\omega_2$ is the unit volume form on $S^2$ and $F_3$ a 3-form in $X_6$. The 5-form $F_5$ is the corresponding field strength associated to the aforementioned D3-branes \cite{Ooguri:2005vr}.\\

On the other hand,  we have learned in the past decade that CY compactifications in string theory yields the presence of flat  potentials in the low energy effective theory, which becomes moduli-dependent once internal fluxes are turned on.  The consequent back-reaction forces the departure of the nice and smooth geometry on CY manifolds into manifolds with generalized geometry. In this context, string compactifications to  Minkowski, anti-de-Sitter (AdS) and de Sitter spaces have been extensively studied in the past few years \cite{Grana:2005jc, Dall'Agata:2001zh, Micu:2007rd, Silverstein:2007ac} (see also \cite{Dibitetto:2010rg, Neupane:2010ya, Parameswaran:2010ec} for recent studies). In particular, the construction of gauged supergravities from type II compactifications on CY manifolds threaded with Ramond-Ramond (RR) and Neveu-Schawrz-Neveu-Schwarz (NS-NS) fluxes has been studied, where the hypermultiplet scalars become charged under the gauge bosons in the vector multiplet.\\ 

However, it is known that not all gauged supergravities are obtained from flux compactifications on CY manifolds, since some of them are constructed by compactifications on manifolds with generalized structures (see, for instance, \cite{Grana:2006hr} and references therein). In other cases, the gauged supergravity does not have a (known) related flux string compactification, although maximal symmetric solutions  have been studied in such scenarios
\cite{Hristov:2009uj}  with the subsequent construction of black-hole solutions \cite{ Hristov:2010eu}. 
Therefore, in the context of gauged supergravities constructed from flux compactifications on manifolds with generalized geometry,  supersymmetric black holes is a topic under recent research. In a similar context, within  the flux compactification scenario (including NS-NS fluxes),
the construction of solutions  of the type $AdS_2\times S^2$ has not been considered in the literature so far.  The objective of this work is to start filling this gap. \\

We study the conditions under which a flux string compactification yields a 4d space-time of the type $AdS_2\times S^2$, on which arbitrary fluxes are present, including NS-NS fluxes.  We find a general constraint that a flux configuration must fulfill in order to be consistent with the desired 4d symmetry. A general solution is difficult to obtain, therefore by constraining the system to have a constant dilaton and in consequence a constant warping factor, we find a simple  minimalist solution. It consists on a compactification of type IIB string theory on a  $SU(3)$-structure manifold,  Ricci-flat  and  with torsion, in the presence of RR and NS-NS 3-form fluxes. We show that our flux configuration is a solution of the Einstein equations with a null contribution to the scalar curvature in 4d and that it satisfies the corresponding  Bianchi identities and integrability conditions. As a result, we provide a novel way to obtain the $AdS_2\times S^2$ geometry as a near-horizon solution. 
  It is also important to mention that the known solution consisting on 5-form RR fluxes is also a solution in our setup.\\

%how extra dimensions become small?: \cite{Brandenberger:1988aj, Durrer:2005nz, Neupane:2010qf} 

% \cite{Giddings:2001yu}.\

Our work is organized as follows. In Sec. 2, we study, in the spirit of \cite{Maldacena:2000mw}, the allowed flux configurations in supergravity compactifications yielding a Ricci-flat space-time formed by the product $AdS_2\times S^2$, by computing the contribution of the energy-momentum tensor of $n$-form fluxes to the scalar curvature.  We also study the compatibility of such configurations with Einstein equations and Bianchi identities. We find a simple minimalist flux configuration fulfilling these constraints consisting on 3-form RR and NS-NS fluxes.
In Sec. 3, we compute the scalar curvature of the 2d $AdS_2$ and $S^2$ spaces as a function of the flux numbers  from the integrability conditions on the 10d spinors on type IIB supergravity. Specifically, we choose the minimalist flux configuration formed by  the 3-forms previously found in Sec. 2. In this case, however, we do not assume any \textit{a priori} relationship among the fluxes. Instead, we look for solutions preserving ${\cal N}=2$ supergravity in 4d. This condition relates the coefficients of the 3-form fluxes, which together with the maximal symmetry assumption for the 2d spaces, imposes the space-time $AdS_2\times S^2$ to be Ricci-flat.  At the end, we briefly comment on nonzero curvature solutions and flux compactifications to 4d with a space-time preserving a black-hole symmetry. Finally we give our conclusions. The appendix is devoted to show our conventions and notations.

%---------------------------------------------------section 2--------------------------------------------------------------------
\section{Flux supergravity compactification}
In this section, we study which type of fluxes are compatible with a $SO(1,1)\times SO(2)\times SU(3)$ symmetry, which corresponds to  a compactification of 10d type IIB string theory on a 6-dimensional manifold $X_6$ with $SU(3)$ structure, into a 4d space-time with a geometry of the type $AdS_2\times S^2$. The allowed fluxes could have different contributions to the scalar 4d curvature if and only if the total contribution vanishes.\\

For that, let us start by considering the most generic 10d metric  compatible with Poincar\'e invariance in 4d  given by \cite{deWit:1986xg, Ortin:2004ms} 
\begin{eqnarray}
ds^2= e^{2A(y)}(\widetilde g_{ij}dx^idx^j + \wt g_{ab}dx^adx^b)+h_{mn}dy^mdy^n.
\label{metric1}
\end{eqnarray} 
The coordinates in $AdS_2$ are labeled by indices $i,j=0,1$ and coordinates in $S^2$ are labeled by indices $a,b=2,3$. For a generic 4d coordinate, we shall use the standard Greek indices $\mu,\nu=0,\cdots ,3$ .\\ 

By the Einstein trace-reversed equations, the Ricci scalar $R(\widetilde{g}_{ij})\equiv \widetilde{R}_{(1)}$ for $AdS_2$ satisfies the relationship,
\begin{eqnarray}
\widetilde{R}_{(1)}+e^{2A}(-T^i_i+\frac{1}{4}T^L_L)=e^{-2A}\nabla^2 e^{2A},
\label{trace}
\end{eqnarray}
where  $T_{MN}$ is the energy-momentum tensor in 10d. Integration over the internal manifold fixes the right-hand side of this equation to vanish, establishing a relation among the curvature of the 4d space-time and  
the field content contribution carried by the second term in the left-hand side. The contribution of a general $n$-form ${\cal F}_n$ is given by the standard expression of its energy-momentum tensor \cite{Maldacena:2000mw, Giddings:2001yu}, which reads
\begin{eqnarray}
{\cal T}_{(1)}\equiv -T^i_i+\frac{1}{4}T^L_L=-{\cal F}_{i M_1 \dots M_{n-1}}{\cal F}^{i M_1\dots M_{n-1}} + \frac{n-1}{4n}{\cal F}^2.
\label{calT}
\end{eqnarray}
A similar result is obtained for $\widetilde{R}_{(2)}$ and ${\cal T}_{(2)}$ for $S^2$.
Hence, to preserve a $SO(1,1)\times SO(2)$ symmetry in 4d, it is necessary to consider
specific flux configurations. Internal fluxes, denoted by ${\cal F}^{int}_n$, with all its legs on the internal manifold and fluxes with two legs on $AdS_2$ or $S^2$, are allowed by this symmetry. The latter are of the form ${\cal F}_n= \omega_2\wedge {\cal F}_{n-2}$ , where $\omega_2$ is a 2-form with coordinates on one of the two subspaces. As we shall see (and opposite to the internal fluxes), the contribution of these $n$-forms ${\cal F}_n$ to the curvature can be  positive or  negative.\\

\subsection{Ricci flat space from supergravity flux compactification}
Following \cite{Maldacena:2000mw, Giddings:2001yu}, we study the contribution to the 4d Ricci scalar $\widetilde{R}$ by fluxes compatible with the symmetry $SO(1,1)\times SO(2)$. For that 
let us firstly consider a   
$n$-form flux ${\cal F}_n=\omega_2\wedge f_{n-2}$,  where $\omega_2$
is the  volume 2-form of the space-time $AdS_2$. In this case, the first term in the right-hand side of Eq.(\ref{calT}) is given by
\begin{eqnarray}
F_{jL_1\dots L_{n-1}}F^{jL_1\dots L_{n-1}}=\frac{2}{n}F^2
\end{eqnarray}
from which the corresponding contribution to $\widetilde{R}$ by ${\cal T}_{(1)}$ is
\begin{eqnarray}
{\cal T}_{(1)}=\frac{n-9}{4n}{\cal F}^2.
\label{That}
\end{eqnarray}
Since ${\cal F}^2\leq 0$ it follows that all field strengths $n$-forms in type IIB theory contribute positively to ${\cal T}_{(1)}$ and negatively to $\widetilde{R}_{(1)}$, except for 9-forms for which the 2d curvature vanishes.\\

Another set of forms compatible with the symmetry consists on fluxes of the form ${\cal G}_n=\wt\omega_2\wedge g_{n-2}$ where $\wt\omega_2$ is the volume 2-form of $S^2$. For these kind of fluxes, the contribution to $\widetilde{R}_{(2)}$ by ${\cal T}_{(2)}$ has the same form as ${\cal T}_{(1)}$,   
\begin{eqnarray}
{\cal T}_{(2)}=\frac{n-9}{4n}G_n^2.
\label{GT}
\end{eqnarray}
Therefore, the contribution to the corresponding curvature of any $n$-form in type IIB is always positive.
On the other hand, internal fluxes ${\cal F}_n$ contribute negatively to $R$ with
${\cal T}= \frac{n-1}{2n}{\cal F}^2_n$,
while fluxes of the form ${\cal G}_n= Vol_4\wedge h_{n-4}$ contribute to the curvature by
${\cal T} = -\frac{9-n}{2n}{\cal G}^2_n$.\\

From this, we see that  a Ricci-flat 4d space-time\footnote{As the reader can easily check, a Ricci-flat space-time is just one possible solution. We can also construct a space-time with a positive or negative Ricci scalar,  as seen in Sec. 3.3.} is a permissible solution from 10d supergravity flux compactification into a 4d space-time given by the 
$AdS_2\times S^2$ geometry since
\begin{eqnarray}
\widetilde{R}_4=\widetilde{R}_{(1)}+\widetilde{R}_{(2)}= -e^{2A}{\cal T}=-e^{2A}({\cal T}_{(1)}+{\cal T}_{(2)})\;,
\end{eqnarray}
vanishes for flux configurations for which ${\cal T}_{(1)}+{\cal T}_{(2)}=0$. It is worth mentioning that this condition can be fulfilled not only by  RR 5-forms of the type $F_5=\omega_2\wedge F_3$, as mentioned at the introduction, but by a wide number of flux configurations.  Notice as well that the fluxes ${\cal F}_n$ and ${\cal G}_n$ can be chosen such that no tadpole is generated in the internal space $X_6$, implying that extra negative-tensioned objects, as orientifolds, are not required. This fact allows to keep two supersymmetries in four dimensions.\\

Before continuing, it is important to keep in mind that our goal is to select a simple flux configuration involving  NS-NS fluxes and to show explicitly that it is compatible with Bianchi identities and with the integrability conditions on the ten-dimensional spinors.

% For the later, we shall compute the corresponding scalar curvatures,  which must equal $\widetilde{R}_{(1)}$ and  $\widetilde{R}_{(2)}$ as computed in this section by the contribution of the energy-momentum tensor.

\subsection{Examples} 
We focus on type IIB supergravity compactifications. As a first example, consider a 5-form $F_5$  of the form $f_2\wedge F_3$ with coefficients given by\footnote{Remember that we are labeling 2d coordinates $x$ on $AdS_2$ with letters $\{i,j,k,l\}$ and on $S^2$ with letters $\{a,b,c,d\}$, while internal coordinate are denoted by indices $\{m,n,p,...\}$.}  $F_{ijmnp}$ and
$F_{abmnp}$. The corresponding 2d scalar curvatures are then given by
\begin{eqnarray}
\widetilde{R}_{(1)}=-\frac{e^{2A(y)}}{5}|F_5|^2, \qquad
\widetilde{R}_{(2)}= \frac{e^{2A(y)}}{5}|F_5|^2.
\end{eqnarray}
Therefore,  the 4d curvature vanishes. Ignoring the back-reaction of such fluxes on the geometry of the internal manifold, this case corresponds to a compactification on a CY to a 4d space-time with a geometry of the type $AdS_2\times S^2$. This is precisely the well-known scenario described in  \cite{Ooguri:2005vr} in which the near-horizon geometry is constructed from a flux compactification on a CY.\\

A more general flux configuration can be chosen.  
Consider, for instance, an $n$-form $H_n$ with two legs on $AdS_2$ (and the rest of them on the internal space $X_6$) and an internal flux ${\cal F}_m$. The corresponding 4d scalar curvature vanishes  if
\begin{eqnarray}
|H^2_n|= \frac{2n}{9-n}\frac{m-1}{m}{\cal F}^2_m.
\end{eqnarray}

Simpler flux configurations can be studied. In particular, we shall focus on the case in which NS-NS 3-form fluxes are turned on. Specifically, let us consider the flux configuration consisting on a NS-NS flux $H_3$ and a RR flux $F_3$ given by
\begin{eqnarray}
H_3&=& \left(Ndx^0\wedge dx^1 + Mdx^2\wedge dx^3\right)\wedge d\alpha,\nn\\
F_3&=& \left(Pdx^0\wedge dx^1 + Q dx^2\wedge dx^3\right)\wedge d\alpha,
\label{fluxesRB}
\end{eqnarray}
with $\alpha$  a function of internal coordinates.  The curvatures, according to  Eqs. (\ref{That}), are given by
\begin{eqnarray}
\widetilde{R}_{(1)}&=& -2e^{2A(y)}(N^2+P^2)(\nabla\alpha)^2, \quad \text{and}\nonumber\\
 \quad \wt{R}_{(2)}&=&2e^{2A(y)}(M^2+Q^2)(\nabla\alpha)^2.
 \label{curvauresE}
\end{eqnarray}
Then, we  see that by taking $M^2+Q^2=N^2+P^2$, the total 4d curvature vanishes. %%%%NOTAR QUE ESTO ES VALIDO PORQUE  LOS TERMINOS DE LA ACCION SON DIFERENTES PARA CADA CAMPO
Under these conditions, the 4d space-time with  an $AdS_2\times S^2$ geometry becomes the near-horizon limit of an extremal Reisnner-Nordstr\"om black-hole.\\

Some comments are given in order: First, notice that we are assuming an internal six-dimensional space with $SU(3)$ structure, for which it is not possible to expand a flux in terms of internal vector components. That is the reason we have assumed a smeared internal leg for the 3-form fluxes as the most general case. Second, 
observe that in this case, $H_3\wedge F_3=0$.  This means that our choice is consistent with the absence of a RR fiveform flux. Also, notice that this does not force the six-dimensional internal space to be a CY manifold, since torsion terms are induced as we shall mention in the next section. Meanwhile, our next step is verify that the above flux  configuration is compatible with 10d Einstein equations and with the corresponding Bianchi identities.

\subsection{Einstein equations}
Following \cite{Giddings:2001yu}, we start by computing the corresponding 2d Ricci tensors. 
Written in the string frame,  the bosonic part of the type IIB superstring action reads
\begin{equation}
S=\frac{1}{2\kappa^2_{10}}\int d^{10}x~\sqrt{-G}e^{-2\phi}\left(R-\frac{1}{12~Im~\tau}|G_3|^2\right),
\end{equation}
with $G_{MN}$ being the 10d metric and $G_3=F_3-\tau H_3$. Then, it follows that the 10-dimensional component of the Ricci tensor is given by,
\begin{eqnarray}
R_{MN}=-\frac{1}{Im~\tau}\left(\frac{G_3^2}{48}G_{MN}-\frac{1}{4}G_{MQR}\bar{G}_N^{\ \ QR}\right).
\label{RMN}
\end{eqnarray}
To preserve the symmetries of the compactification setup, the most general metric we shall consider is 
\begin{eqnarray}
ds^2=e^{2A(y)}\wt{g}_{ij}dx^idx^j + e^{2B(y)}\wt{g}_{ab}dx^adx^b  + e^{-2A(y)}\wt{h}_{mn} dy^mdy^n.
\end{eqnarray}
Here, we have assumed in principle two different warping factors for the two-dimensional subspaces $AdS_2$ and $S^2$.  The corresponding Ricci tensors are
\begin{eqnarray}
R_{ij}= \wt{R}_{ij}-e^{4A}\left(\wt\nabla^2A+2 \wt\nabla A\cdot\wt\nabla B - 2(\wt\nabla A)^2\right)\wt{g}_{ij},
\end{eqnarray}
\begin{eqnarray}
R_{ab}= \wt{R}_{ab}-e^{2(A+B)}\left(\wt\nabla^2B-2 \wt\nabla A\cdot\wt\nabla B + 2(\wt\nabla B)^2\right)\wt{g}_{ab},
\end{eqnarray}
where, following standard notation, $\tilde\nabla$ is the covariant derivative with respect to the nonwarped metric $\wt{h}_{mn}$. 
From these expressions and the corresponding components of Eq.(\ref{RMN}),  the Ricci tensors for each subspace written in terms of the nonwarped metric are given by
\begin{eqnarray}
\wt{R}_{ij}&=& e^{4A}\left(\widetilde\nabla^2A+2 \wt\nabla A\cdot\wt\nabla B - 2(\wt\nabla A)^2\right)\wt{g}_{ij}-\frac{1}{Im~\tau}\left(\frac{G_3^2}{48}G_{ij}-\frac{1}{4}G_{iQR}\bar{G}_j^{\ \ QR}\right),\nn\\
\wt{R}_{ab}&=&e^{2(A+B)}\left(\widetilde\nabla^2B-2 \wt\nabla A\cdot\wt\nabla B + 2(\wt\nabla B)^2\right)\wt{g}_{ab}\nn\\
&&-\frac{1}{Im~\tau}\left(\frac{G_3^2}{48}G_{ab}-\frac{1}{4}G_{aQR}\bar{G}_b^{\ \ QR}\right).
\label{einsteins}
\end{eqnarray}
As it was done for the four-dimensional maximally symmetric case  \cite{Giddings:2001yu}, we proceed to compute the corresponding Bianchi identities and look for possible  constraints the fluxes must fulfill. This method has been used to determine the supersymmetric conditions on the fluxes for type IIA and type IIB compactifications on maximally symmetric 4d spaces. We shall not give a rigorous proof that such method works for our case; instead, we shall show that such method consistently fixes some variables for the special case we are studying.\\

\subsection{Bianchi identities}
Since we are dealing with 3-form fluxes with a smeared leg on the internal space, the Bianchi identities for $F_3$ and $H_3$ are trivially satisfied. To look for more stringent conditions on the fluxes, we study the dual Bianchi identities
\begin{equation}
d\ast F_3=d\ast H_3=0,
\end{equation}
which yields the pair of equations 
\begin{eqnarray}
(Q+\tau M)\left(-2\partial_m(2A+B)\tilde{\partial}^m\alpha+\tilde{\partial}^2\alpha\right)&=&0\nonumber\\
(P+\tau N)\left(+2\partial_m( -4A+B)\tilde{\partial}^m\alpha+\tilde{\partial}^2\alpha\right)&=&0.
\end{eqnarray}
It immediately follows that $A=B$ for nonvanishing   $Q+\tau M$ and $P+\tau N$. It seems that, even though the two-dimensional subspaces $AdS_2$ and $S^2$ are independent, they share the same warping factor in a background threaded with fluxes given by (\ref{fluxesRB}). Hence, under this condition, the above pair of equations reduces to
\begin{eqnarray}
\tilde{\partial}^2\alpha=6e^{-2A}\partial_mA\partial^m\alpha=\frac{3}{2}e^{-6A}(\partial_me^{4A})(\partial^m\alpha).
\end{eqnarray}

Comparing Einstein Eqs.(\ref{einsteins}) with the Bianchi identity for components in $AdS_2$, we see that 
\begin{eqnarray}
\tilde{\nabla}^2(e^{4A}-\alpha)&=&2\tilde{R}_{(1)}+\frac{1}{2}e^{-6A}(\partial_me^{4A}\partial^m e^{4A})-\frac{3}{2}\partial_me^{4A}\partial^m\alpha\nonumber\\
&&+\frac{1}{4Im~\tau}\left(-P^2-\tau\bar\tau N^2+2 (Im~\tau)PN\right)e^{-2A}\partial_m\alpha\partial^m\alpha,
\label{m2}
\end{eqnarray}
where we have  taken  the configuration in  (\ref{fluxesRB}) to compute the flux components. Similarly, for coordinates on $S^2$, we get
\begin{eqnarray}
\tilde{\nabla}^2(e^{4A}-\alpha)=2\tilde{R}_{(2)}+\frac{1}{2}e^{-6A}(\partial_me^{4A}\partial^m e^{4A})-\frac{3}{2}\partial_me^{4A}\partial^m\alpha\nonumber\\
+\frac{1}{4Im~\tau}\left(Q^2+\tau\bar\tau M^2+2 (Im~\tau)QM\right)e^{-2A}\partial_m\alpha\partial^m\alpha.
\label{m22}
\end{eqnarray}
Since $\wt{R}_4=\wt{R}_{(1)}+\wt{R}_{(2)}$ must vanish for $AdS_2\times S^2$,  adding Eqs. (\ref{m2}) and (\ref{m22}) yields
\begin{eqnarray}
\tilde{\nabla}^2(e^{4A}-\alpha)&=&e^{-6A}(\partial_me^{4A}\partial^m e^{4A})-3\partial_me^{4A}\partial^m\alpha+\frac{1}{4Im~\tau}\left((Q^2-P^2)-\tau\bar\tau (M^2-N^2)\right.\nn\\
&&\left.+2 (Im~\tau)(QM+PN)\right)e^{-2A}\partial_m\alpha\partial^m\alpha.
\label{compact}
\end{eqnarray}
The left-hand side of this equation integrates to zero in a compact manifold constraining the values of the flux numbers and the warping factor $A$.\\

 Solutions with a nonconstant warping factor seem difficult to find in the general case. The most general solution would relate the warping factor $A$ with the function $\alpha$ and the flux numbers. Such analysis is beyond the scope of this study and we leave it for a future work. \\
 
Instead, we concentrate on the simplest
solution involving a constant warping factor  $A$. Our goal is to find the minimal conditions under which we can construct the Robinson-Bertotti solution on four dimensions.  There are some cases to consider (all of them satisfying the general condition  for $\hat{\cal T}=0$,  $M^2+Q^2=N^2+P^2$) in which Eq. (\ref{compact}) vanishes:

\begin{enumerate}
\item
$(M,N,P,Q)\neq 0$  and $M=-N, Q=P$
\item
$P=0$  and $\frac{M}{Q}=\frac{\tau\bar\tau -1}{2~Im~\tau}$
\item
$Q=0$ and $ \frac{N}{P}=\frac{\tau\bar\tau +1}{2~Im~\tau}$
\item
$N=0$ and $ \frac{Q}{M}=\frac{\tau\bar\tau -1}{2~Im~\tau}$
\item
$M=0$ and $ \frac{P}{N}=\frac{\tau\bar\tau +1}{2~Im~\tau}$
\end{enumerate}

As previously mentioned, all the above cases satisfy the constraint $H_3\wedge F_3=0$ as expected for a configuration without a 5-form RR flux.\\

By choosing one of these constraints, we proceed to compute the curvature scalars from the integrability conditions on the 2d components of 10d spinors. It turns out that conditions (2) to (4) are easier to be considered in this procedure. Therefore, we shall concentrate on a configuration of RR and NS-NS 3-form fluxes given by (\ref{fluxesRB}) satisfying one of these constraints.\\

%-----------------------------------------------------section 3---------------------------------------------------------------------------
\section{Near-horizon geometry from RR and NS-NS 3-form fluxes}
Our main goal in this section is to compute the scalar curvature of the 2d spaces $AdS_2$ and $S^2$ (as a function  of a specific flux configuration) from the integrability conditions on the 2d components of  10d spinors  within a type IIB supergravity flux compactification scenario. For that, we assume a 10d space-time of the form  $AdS_2\times S^2\times X_6$, where the 2d spaces have arbitrary curvatures and $X_6$ is assumed to have a $SU(3)$ structure. We focus exclusively on the flux configuration formed by 3-form NS-NS and RR fluxes found in the previous section. In principle, we do not assume any relationship among the flux coefficients, but instead we look for solutions preserving  ${\cal N}=2$ 4d supergravity. These conditions restrict the $AdS_2$ and $S^2$ curvatures to be equal in magnitude, rendering the 4d space-time to be Ricci-flat.

\subsection{Ricci curvature from integrability conditions}

Let us consider the variation of fermi fields in  type IIB supergravity in the presence of 3-form fluxes described by Eq. (\ref{fluxesRB}). When supersymmetry is preserved, the gravitino variation reads
\begin{eqnarray}
\delta\Psi_M=\nabla_M\epsilon-\frac{1}{4}\not{\!\!H}_M\sigma^3\epsilon+\frac{1}{16}e^\phi \not{\!\!F}_3\Gamma_M\sigma^1\epsilon=0
\label{gravitino}
\end{eqnarray}
with $\sigma$ being the Pauli matrices, and  $\epsilon=\left(\begin{array}{c}
\epsilon^1\\
\epsilon^2
\end{array}\right)$ the chiral spinors. We can observe that the ten-dimensional chiral spinors $\epsilon^{1,2}$ are correlated by the vanishing of the gravitino variation.\\
 
Since we want to compactify a 10d supergravity theory into a maximally supersymmetric space which admits two maximal symmetric subspaces, we split the structure group of the
tangent bundle $SO(1,9)$ into $SO(1,1) \times SO(2) \times SU(3)$,
where we have chosen a 6d internal space of $SU(3)$
structure. Using the vanishing supersymmetric variation of the fermi fields, we shall obtain  an independent equation for each of the two 10d spinors $\epsilon^{1,2}$.\\

% Accordingly we split the 10d spinors as
%%
%\begin{eqnarray}
%  \label{split}
%  \epsilon^1&=& \zeta_+^1\otimes \eta^1_++\zeta_-^1\otimes \eta^1_-=a^1 \theta^1 \otimes \xi^1 \otimes \eta^1_+ + 
%  b^1 \bar\theta^1 \otimes \xi^1 \otimes  \eta^1_- \; ,\nn\\
%   \epsilon^2&=& \zeta_+^2\otimes \eta^2_++\zeta_-^2\otimes \eta^2_-=a^2 \theta^2 \otimes \xi^2 \otimes \eta^2_+ + 
%  b^2 \bar\theta^2 \otimes \xi^2 \otimes  \eta^2_- \; ,
% \end{eqnarray}
%%
%where $\theta$ is a Weyl spinor on the $AdS_2$ space-time, $\xi$ is a
%Weyl spinor on $S^2$ and $\eta$ denotes the invariant spinor on the
%manifold with $SU(3)$ structure. The 4d spinors are denoted as $\zeta_\pm$. For ten-dimensional Majorana-Weyl spinors
%we need $a=b$.\\

%Recall that our goal is  to preserve two supersymmetries in the effective 4d theory, therefore there must  not be a relation between the 4d spinors $\zeta^{1,2}$.  

Let us start by taking the 2d component of $\delta\Psi^1$ in a background threaded with the flux content (\ref{fluxesRB}).  Using the fact that
\begin{eqnarray}
[\Gamma_i,\Gamma_{jkm}]=\{\Gamma_i, \Gamma_{abm}\}=0, 
\end{eqnarray}
the $M_2$-component of the gravitino variation is written as
\begin{eqnarray}
\nabla_i\epsilon-\frac{1}{4}\not{\!\!H}_i\sigma^3\epsilon+\frac{1}{16}e^\phi \Gamma_i(\not{\!\!F^{(1)}}_3-\not{\!\!F^{(2)}}_3) \sigma^1\epsilon=0,
\end{eqnarray}
where $\not{\!\!F^{(1)}}_3= F^{01m}\Gamma_{01m}$ and $\not{\!\!F^{(2)}}_3= F^{23m}\Gamma_{23m}$. Our strategy here consists on commuting $\Gamma_i$ with $\not{\!\!F}_3$ in the last term of the above equation by using the dilatino variation for a constant dilaton given by
\begin{eqnarray}
\delta\lambda=-\frac{1}{2}\not{\!\!H}_3\sigma^3\epsilon-\frac{1}{4}e^{\phi}\not{\!\!F}_3\sigma^1\epsilon=0,
\end{eqnarray}
to decouple the ten-dimensional spinors $\epsilon^{1,2}$. However, this seems difficult  to perform unless $\not{\!\!F^{(1)}}_3$ or $\not{\!\!F^{(2)}}_3$ vanishes, but according to section 3, this is an available condition on the fluxes. \\

In consequence, we shall consider the case in which $P=0$, which corresponds to $F^{01m}=0$. Therefore, the $i$-component of the gravitino variation reads,
\begin{eqnarray}
\left(\nabla_i-\frac{1}{4}\H_i+\frac{1}{8}\Gamma_i\H_3\right)\epsilon^1=0.\label{nabla1},
\end{eqnarray}
with a similar expression for the $a$-component.\\

It is important to notice that the decoupling of the two 10d spinors is possible only in the presence of nontrivial fluxes $H_3$ and $F_3$. From now on, we shall concentrate on the equation involving $\epsilon^1$. A similar analysis is performed on the second spinor $\epsilon^2$ with similar results. We comment on those results at end of this section.\\

Equation (\ref{nabla1}) can be expressed as 
\begin{eqnarray}
\nabla^{(T)}_i\epsilon^1=(\nabla_i+\kappa_i)\epsilon^1=0,
\end{eqnarray}
with $\kappa_i=-\frac{1}{4}\H_i+\frac{1}{8}\Gamma_i{\H}_3$. The chiral ten-dimensional spinor is not covariantly constant under the Levi-Civita connection, but under $\nabla^{(T)}$.\\

The second step consists on computing the corresponding components of the connection directly from the metric we are working with.  In this case, the metric is given by
$ds^2=e^{2A(y)}\wt{g}_{\mu\nu}dx^\mu dx^\nu+ h_{mn}dy^mdy^n$, from which the $i$-component of the covariant derivative of the spinor $\epsilon^1$ is
\begin{eqnarray}
\nabla^T_i\epsilon^1=\left(\widetilde\nabla_i-\frac{1}{2}\gamma_i\widetilde\gamma\otimes\widetilde\sigma\otimes\not{\!\partial}A+\kappa_i\right)\epsilon^1=0,
\end{eqnarray}
where we have denoted the covariant derivative with respect to the  metric $g_{ij}$ as $\widetilde{\nabla}$ and $\nabla_i=\widetilde{\nabla}_i-\frac{1}{2}\gamma_i\widetilde\gamma\otimes\widetilde\sigma\otimes\not{\!\partial}A$.\\
From the integrability condition on the connection $\wt\nabla$,  we have that 
\begin{equation}
[\widetilde\nabla_i,\widetilde\nabla_j]\epsilon^1=\frac{1}{4}\widetilde{R}_{ij}^{\ \ \ kl}\gamma_{kl} \epsilon^1,
\end{equation}
and from the gravitino variation we also have that
\begin{eqnarray}
[\cov_i, \cov_j]\epsilon^1 = -(\partial_nA\partial^nA)\gamma_{ij} +\frac{1}{2}[\gamma_i\widetilde\gamma\otimes\widetilde\sigma\otimes\not{\!\partial}A, \kappa_j]+\frac{1}{2}[\kappa_i,\gamma_j\widetilde\gamma\otimes\widetilde\sigma\otimes\not{\!\partial}A]+[\kappa_i,\kappa_j]\epsilon^1.\nonumber\\
\end{eqnarray}
Let us here consider the simple case previously mentioned,  by taking a constant warping factor $A$ . In that case,  the Riemann tensor is given by
\begin{eqnarray}
\frac{1}{4}\widetilde{R}_{ij}^{\ \ \ kl}\gamma_{kl} -[\kappa_i, \kappa_j]=0.
\label{Riemann}
\end{eqnarray}
Notice that in the fluxless case, the contorsion and, therefore, the  Riemann tensor vanishes for a constant warping factor , leading  to a Minkowskian space-time . In this case, however, the contorsion brings an extra term even for a constant warping factor. Then, we have that
\begin{eqnarray}
[\kappa_0,\kappa_1]&=&\frac{1}{16} [\not{\!\!H}_0, \not{\!\!H}_1]
+\frac{1}{64} [\Gamma_0\not{\!\!H}_3, \Gamma_1\not{\!\!H}_3]
 -\frac{1}{32}[\not{\!\!H}_0, \Gamma_1\not{\!\!H}_3]
 -\frac{1}{32}[\Gamma_0\not{\!\!H}_3, \not{\!\!H}_1]\nn\\
 &=& -\frac{1}{32}(N^2+M^2)(\nabla\alpha^2)\gamma_{01}.
\end{eqnarray}
Therefore from Eq.(\ref{Riemann})  and by the maximal symmetry on $AdS_2$ with symmetry $SO(1,1)$, the corresponding 2d Ricci scalar is given by
\begin{eqnarray}
R_{(1)}=-\frac{1}{8}(N^2+M^2)(\nabla\alpha)^2.
\end{eqnarray}
Similarly, for $S^2$ with symmetry $SO(2)$, the scalar curvature is $\wt{R}_{(2)}=-\wt{R}_{(1)}$.\\

Hence, there is a  unique 4d  solution of this system, namely, the near-horizon geometry $AdS_2\times S^2$ with $\wt{R}_4=0$. Contrary to the analysis in Sec. 2, where a relation among fluxes must be taken by hand in order to obtain the near-horizon geometry, here the relation among fluxes is established by requiring ${\cal N}=2$ supergravity in 4d (which implies the decoupling of the spinors $\epsilon^1$ and $\epsilon^2$ in 
(\ref{gravitino})).\\

Even more, notice that we have found an alternative way for constructing a 4d space-time with an $AdS_2\times S^2$ symmetry by turning on 3-form fluxes, including NS-NS, rather than the inclusion of only 5-form fluxes. We conclude, therefore, that $AdS_2\times S^2$ geometry is not necessarily constructed as the limit of an extreme black-hole, but also by a different choice on the internal manifold (see Sec. 3.2) and on the type of fluxes we turned on.\\

Let us emphasize some interesting facts: 
\begin{enumerate}
\item For $M=N=0$, i.e., in the fluxless case, both curvatures vanish and we recover the Minkowski 4d space-time. 
\item Although it seems that RR fluxes do not play a role in the curvature, they can not vanish,  otherwise  the contribution to $\wt{R}_4$ from Einstein equations by ${\cal T}$ would not be zero.
\item From the curvatures $\wt{R}_{(1)}$ and $\wt{R}_{(2)}$, it is possible to construct an effective 4d metric of the form
\begin{eqnarray}
ds^2_4= -\frac{x_1^2}{h}dx_0^2+\frac{h}{x_1^2}dx_1^2+hdx_2^2+h~sin^2x_2dx_3^2\:,
\end{eqnarray}
with $h= 2/|\wt{R}_{(1)}|$. Besides reproducing the above curvatures, this metric must be a solution of the effective field theory in 4d. By ignoring the presence of moduli scalar fields produced by the compactification procedure, the effective theory contains gravity and fields with two antisymmetric indices induced by the presence of RR and NS-NS fields, which are effectively interpreted as electromagnetic tensor fields with no sources. Notice that the NS-NS and RR fluxes have both a smeared leg on the internal manifold.\\

Therefore, in the presence of an homogenous electromagnetic field of the form
 \begin{eqnarray}
F_{t\rho}= 2|\wt{R}_{(1)}|=\frac{1}{4}(N^2+M^2)(\nabla\alpha)^2,
\end{eqnarray}
there is a unique solution of the effective Einstein-Maxwell equations.
 This is the near-horizon metric of an extremal black hole, known as Robinson-Bertotti solution. Notice that coordinates of the $H_3$ and $F_3$ fluxes are effectively related to the homogenous electromagnetic field. It would be interesting to construct the effective 4d gauged supergravity by a compactification on a Ricci-flat internal manifold with torsion, and recover the above field as a function of  the NS-NS and RR fluxes.
 \item The curvature of each subspace is proportional to the flux number $(N^2+M^2)$.  A big flux number corresponds to a highly curved 2d subspaces and to a smaller horizon area.
\item  Observe that the integrability conditions on the second 10d spinor lead to exactly the same result, since for that case
\begin{equation}
\kappa_i=\frac{1}{4}\not{\!\!H}_i-\frac{1}{8}\Gamma_i\not{\!\!H},
\end{equation}
and both 2d  curvatures are not modified  with respect to the curvatures computed by the spinor $\epsilon^1$ equations.
\item A different case would consider  a RR flux configuration in which $Q=0$. In such a case it is possible to show that a solution of the type $AdS_2\times S^2$ as near-horizon geometries is also obtained. 
\end{enumerate}

%---------------------------------------SUBSECTION
\subsection{Internal manifold and its torsion}
At this point, we have some glimpses about the geometry of the internal manifold corresponding to the simplest case we have analyzed, where the warping factor is constant and the flux configuration is given by (\ref{fluxesRB}). Although a detailed classification of the internal geometry leading to  ${\cal N}=2$ supersymmetry  in 4d, consisting on two maximal symmetric subspaces,  is beyond the scope of this work, we can at least say some generalities.\\

The internal spinors satisfy the equation
$\nabla^{(T)}_m\eta_\pm =0$,
this is,  $\nabla_m\eta=\kappa_m\eta$, where   the contorsion $\kappa$ has an intrinsic part $\kappa^0$ which is an element of $\lambda^1\otimes su(3)^\perp$ (the component in $su(3)$ acts trivially on the spinors) \cite{Louis:2002ny, Grana:2004bg} . This intrinsic torsion can be decomposed into $SU(3)$ representations, denoted ${\cal W}_1, {\cal W}_2, {\cal W}_3, {\cal W}_4, {\cal W}_5$ (see Appendix for notation).\\

In our case, the torsion component of the derivative operator, is given by the direct product of gamma matrices acting on the internal spinor. Then, for the K\"ahler form we have that
\begin{eqnarray}
-\kappa_qJ_{mn}=\left(\frac{1}{4}\not{\!\!H}_m-\frac{1}{16}e^\phi\not{\!\!F}_3\Gamma_m\right)\eta^\dag\gamma_{mn}\eta\;.
\end{eqnarray}
By turning on fields with legs on extended coordinates we get that , $\not{\!\!H}_m= (\tilde{\gamma}\otimes\mathbf{1}\otimes\mathbf{1})N\nabla_m\alpha+(\tilde{\gamma}\otimes\tilde\sigma\otimes\mathbf{1})M\nabla_m\alpha$. This implies that $dJ\sim \kappa_q J_{mn}(dz^q+d\bar{z}^q)\wedge dz^m\wedge d\bar{z}^n$, i.e., it is in the $(3\oplus 3)+(3\oplus 3)$  of $SU(3)$. Then the torsion representation ${\cal W}_1$ corresponding to $(3,0)$ and $(0,3)$-forms is absent. For the general case, in which internal fluxes are turned on, it is possible to get torsion ${\cal W}_1$ representations.\\

For the covariant derivative of the holomorphic 3-form, we get that $d\Omega_3$ can be a $(3,1)$ or $(1,3)$ forms, but only 9 out of 18 different  $(2,2)$ forms, for which we have  all ${\cal W}_5$ $SU(3)$ representations, but a half of ${\cal W}_1$ and ${\cal W}_2$.  We expect, as well, that for the generic case all torsion representations ${\cal W}_1$, ${\cal W}_2$, and ${\cal W}_5$ would be in principle allowed.\\

For the simplest case we have previously concentrated on, we see that the internal curvature vanishes. This follows straightforward from the fact that an $n$-form contributes to the internal curvature as \cite{Douglas:2010rt}
\begin{eqnarray}
R_{X_6}= T^m_m-\frac{3}{4}T^L_L= \frac{3}{4}\left(\frac{9-n}{n}\right)F^2_n\;,
\end{eqnarray}
for which  the flux configuration given in (\ref{fluxesRB}) reduces to
\begin{eqnarray}
R_{X_6}=\frac{3}{2}(M^2-N^2+Q^2)\;,
\end{eqnarray}
where the curvature is taken with respect to the Riemann connection. Hence, the internal curvature vanishes for $N^2=M^2+Q^2$ as required from ${\cal T}=0$. This implies that the fluxes do not contribute to internal energy, rendering the internal manifold to be Ricci flat, with an $SU(3)$ structure (by assumption) and with a constant warping factor. In the fluxless case, this forces the manifold to be CY.  In our case, the torsion components are not trivial. The internal manifold must be Ricci-flat with nontrivial torsion components. Recently it was proved in Ref.(\cite{Bedulli:2006ol}) that the scalar curvature of the metric,  induced by the $SU(3)$-structure is expressed in terms of the torsion forms, opening up the possibility to construct manifold with the above mentioned properties. A detailed analysis of the torsion components are beyond the scope of this note, and we left it for future work. \\

%----------------------------------------------------------section 4---------------------------------------------------------------
%\section{Positive scalar curvature from fluxes in a non-maximally symmetric  4d space-time.}

%In this section we briefly comment on the possibility to turn on  flux configurations while  preserving different (non-maximal) symmetries in a 4d space-time and yielding to a positive 4d scalar curvature. Clearly for this to happen, we break explicitly the Poincar\'e invariance in the 4d effective theory. Specifically we shall consider a 4d space-time with i) a  $AdS_2\times S^2$ and ii) a supersymmetric black-hole symmetry. \\

\subsection{Nonzero scalar curvature for ${AdS_2}\times{S}^2$}
In Sec. 2, we studied a particular case in which the configuration of 3-form fluxes (\ref{fluxesRB}) $-$preserving the symmetry of $AdS_2\times S^2$$-$ leads to a 4d space-time with curvature
\begin{equation}
\wt{R}_4= e^{2A(y)}(M^2+N^2-P^2-Q^2),
\end{equation}
which vanishes for the special case in which $M^2+N^2=P^2+Q^2$. A positive or negative curvature of this 4d space-time can be accomplished by taking fluxes which do not fulfill the above equality. Effectively, this corresponds to take different radii on the $AdS_2$ and $ S^2$ factors in the 4d metric. However, from the flux compactification point of view, there are many other different flux configurations yielding to a positive (or negative) 4d curvature rather than taking different values for the flux numbers $N^2+M^2$ and $P^2+Q^2$.\\

The more general flux configuration consisting on a n-form ${\cal F}_n$ with two legs on $AdS_2$, an $m$-form ${\cal G}_m$ with two legs on $S^2$, and an internal  $p$-form flux  ${\cal H}_p$ (all its legs on the internal space $X_6$) yields a 4d curvature given by 
\begin{equation}
\wt{R}_4=e^{2A(y)}( -\frac{9-n}{4n}|{\cal F}^2_n| +\frac{9-m}{4m}{\cal G}^2_m+\frac{1-p}{2p}{\cal H}^2_p).
\end{equation}
Different values for the fluxes yields to positive, negative, or null 4d curvatures.
Observe that, although in these examples the space-time can have a positive Ricci-curvature, it is not asymptotically De-Sitter. However, it is worth noticing that by not requesting a maximally symmetric 4d space-time, the possibility to achieve different values for the curvature increases. Thus, it would be interesting to consider different 4d symmetries in order to look for richer scenarios in which 4d scalar curvature acquire any possible value. One of them involves a 4d black-hole symmetry.\\

\subsection{Black-hole  symmetry}
Consider a type IIB supergravity flux compactification into a 4d space-time with a  metric
\begin{eqnarray}
ds^2_4=e^{2A(y)}\widetilde{g}_{\mu\nu}=e^{2A(y)}\left(-e^{2U(r)}dt^2+e^{-2U(r)}d\vec{x}^2\right).
\end{eqnarray}
The metric $\wt{g}_{\mu\nu}$  describes the 4d space-time around  a single-centered static and supersymmetric black hole \cite{Andrianopoli:2006ub}.  The interplay between fluxes and black holes has established very interesting lines of study, as the stability of black-holes in the presence of fluxes \cite{Danielsson:2006jg}, the attractor mechanism \cite{Kallosh:2005ax, Denef:2000nb} and the construction of black holes in a background threaded with fluxes \cite{Hsu:2006vw}, where the construction of black holes by D3-branes implies the presence of threeform fluxes and a fiveform.\\

Here, we are interested in the minimal set of fluxes which preserve the above symmetry. Notice that we are not constructing a black hole, but just studying what kind of fluxes can be turned on in a supergravity compactification such that they are compatible with the 4d black-hole-like symmetry. One possibility is to consider fluxes which have only one leg on a timelike coordinate or three of them in the three spacelike coordinates.  Following Sec. 2, we want to compute their contribution to the 4d curvature.  \\

The flux contribution to $\wt{R}_4$  by fluxes of the form  $F_n\sim f_{0m_1\cdots m_{n-1}}dx^0\wedge dy^{m_1}\wedge\cdots\wedge dy^{m_{n-1}}$ is 
\begin{eqnarray}
{\cal T}_1=\frac{n-9}{8n}F_n^2,
\end{eqnarray}
while for fluxes of the form  $G_p\sim G_{ijk}g_{m_1\cdots m_{p-3}}dx^i\wedge dx^j\wedge dx^k\wedge dy^{m_1}\wedge\cdots\wedge dy^{m_{p-3}}$, the contribution to $\wt{R}_4$  is
\begin{eqnarray}
{\cal T}_2= \frac{3(p-9)}{8p}G_p^2,
\end{eqnarray}
where we have denoted by $x$ the coordinates on the extended 4d space-time and by $y$ the coordinates on the internal 6d space. Notice that internal $m$-form fluxes ${\cal F}_m$ are still allowed by the symmetry with the usual negative contribution.
Therefore, the 4d scalar curvature $\wt{R}_4$ has 3 different contributions, 
\begin{equation}
\wt{R}_4=-e^{4A(y)}\left(-\frac{n-9}{8n}|F_n^2|+\frac{3(p-9)}{8p}G_p^2+\frac{m-1}{2m}{\cal F}_m^2\right).
\end{equation}
We see that the contribution of fluxes with a timelike leg is negative, rendering the 4d curvature to be positive, negative or null.\\

The existence of a black-hole symmetry can be justified in a qualitative and speculative way as follows. The symmetry we associate to our 4d space-time could be determined in a higher scale than the scale of compactification ${\cal M}_{\text{comp}}$ (which is assumed much larger than the supersymmetry breaking scale). In that situation, primordial supersymmetric black-holes (formed at a scale of energy ${\cal M}_{\text{BH}}$ such that ${\cal M}_{\text{Planck}}<< {\cal M}_{\text{BH}}<{\cal M}_{\text{comp}}$) could affect the symmetry of our space-time.  Before compactification, the ten-dimensional space-time would be asymptotically flat, but it would turn positively-curved after compactification. The curvature would be determined by the fluxes present in the initial configuration. Clearly, a more extensive and detailed study is needed here.\\

On the other hand, an interesting thing to do, concerns the construction of the effective 4d gauged supergravity constructed from the ten-dimensional setup we have studied. Recently, it was shown that gauged 4d supergravities admit flat and negative curved solutions. In particular, an $AdS_2\times S^2$ geometry was found as a solution of a  gauged ${\cal N}=2$ 4d supergravity \cite{Hristov:2009uj}. Even more, the theory contains black-holes solutions as shown in \cite{Hristov:2010eu}. It would be interesting to study the effective gauged supergravity derived from  a general flux configuration and compare it with the solutions shown in those references.

%-----------------------------------------------------------------------------------------CONCLUSIONS---------------------------------------------------------------------------------------------------------------------------------

\section{Conclusions and final comments}
By considering a  type IIB supergravity flux compactification on a six-dimensional manifold with $SU(3)$ structure, 
we study the required conditions  that an arbitrary flux configuration must satisfy to obtain a warped 4d space-time of the type $AdS_2\times S^2$  as a near-horizon geometry. We take into account the possibility to turn on NS-NS fluxes.\\

Among all possible flux configurations,  we concentrate on a simple  minimalist solution consisting only on  RR and NS-NS 3-form fluxes. Out of their three legs, two are on $AdS_2$ or $S^2$, and one is smeared  on the internal manifold.  These fluxes thread a space-time with a constant warping factor. We show that this flux configuration is a solution of the Einstein equations and the corresponding  Bianchi identities. They contribute with a null scalar curvature in 4d  for a specific relationship among the flux coefficients. This renders the $AdS_2\times S^2$ geometry as a near-horizon geometry in 4d.\\

We also compute the  scalar curvature of the 2d spaces $AdS_2$ and $S^2$,  as  function  of the fluxes, from the integrability conditions on the 2d components of  10d spinors. The setup consists on a flux compactification of type IIB supergravity threaded with the 3-form fluxes previously considered.
However, here we do not assume any relationship among the flux coefficients, but instead we look for solutions preserving  ${\cal N}=2$ 4d supergravity. This condition restricts the space-time $AdS_2\times S^2$  to be Ricci-flat.\\

We  comment on some characteristics the internal manifold has, as the fact that it is Ricci-flat with nontrivial torsion components. In the fluxless case, there is no torsion, and the manifold is CY.  A detailed analysis of the torsion components are beyond the scope of this note, and we left it for future work. \\

Summarizing, our work shows a way to construct a 4d space-time with an $AdS_2\times S^2$ symmetry by turning on 3-form fluxes $-$including NS-NS fluxes$-$ as an alternative to the well-known case of five forms associated to D3-branes and considered in literature so far.   Therefore, we conclude  that solutions of the type $AdS_2\times S^2$ as near-horizon geometry are not uniquely constructed as the limit of extreme black-holes in ungauged supergravities, but also by compactifications on internal manifolds with torsion  derived  by the presence of arbitrary flux configurations.   This opens up the possibility to construct black-hole solutions in the context of gauged supergravity in which the near-horizon limit  would be described by our solution. We do not focus on this topic in our present work, but is worth mentioning a recent related study \cite{Kimura:2011sy}.\\

We also study the possibility to obtain different nonzero  values for the curvature of the space-time $AdS_2\times S^2$ by considering an arbitrary flux configuration. It is important to notice that the possibility to achieve different values for the curvature increases
by not requesting a maximally symmetric 4d space-time. Under this perspective, we also study a flux compactification on a 4d space-time with a black-hole symmetry, i.e., with a symmetry derived by the presence of a supersymmetric black hole. We find that the corresponding scalar curvature derived from the contribution of the fluxes to the energy-momentum tensor acquires all possible values. A positive curvature is accomplished by considering a flux with 3 legs on spacial-like coordinates.\\

Recently, it was shown that gauged 4d supergravities admit flat and negative curved solutions. In particular, an $AdS_2\times S^2$ geometry was found as a solution of a  gauged ${\cal N}=2$ 4d supergravity \cite{Hristov:2009uj}. Even more, the theory contains black-holes solutions as shown in \cite{Hristov:2010eu}. It would be interesting to study the effective gauged supergravity derived from  our flux configuration and compare it with the solutions shown in these references.\\

\begin{center}
{\bf Acknowledgements}
\end{center}
We thank A. Micu for collaboration at the beginning of this project and for useful suggestions. We also thank S. Ramos-S\'anchez for proof-reading and for many interesting discussions.
O.L.-B. was partially supported by PROMEP. L.V. is supported by a CONACyT doctoral fellowship.

%-----------------------------------------------------------APPENDIX-----------------------------------------------------------
\appendix
\section{Notation, Conventions and Calculations}

Here, we summarize the conventions and notations we use through the body of this paper. Also, we show some gamma matrix algebra calculations we refer to in the paper.

\subsection{Notations and Conventions}
For the gamma matrices in any dimension, we use 
\begin{eqnarray}
\Gamma_{M_1\cdots M_N}=\frac{1}{N!}\Gamma_{[M_1}\Gamma_{M_2}\cdots\Gamma_{M_N]},
\end{eqnarray}
and for the covariant derivative on a spinor we take the spinor connection as
\begin{eqnarray}
\nabla_M\Psi=\partial_M\Psi-\frac{1}{4}{\omega^{AB}}_M\Gamma_{AB}.
\end{eqnarray}
We are working on a ten-dimensional space-time split in two spaces $M_4\times X_6$. We use Greek index $\mu,\nu,\dots$ to label 4d coordinates and Latin letters $m,n,p,\dots$ for the internal coordinates. Since we also deal with a 4d space-time split into two maximal symmetric spaces, i.e., $M_4=AdS_2\times S^2$, we use Latin letters $i,j,k,l$ to label coordinates on $AdS_2$ with a metric signature $(-,+)$ and letters $a,b,c,d$ for coordinates on $S^2$ with metric signature $(+,+)$.\\

The gamma matrices in ten dimensions can be constructed from lower-dimensional
ones as follows. First note that in ten dimensions (i.e.  for $SO(1,9)$) the
gamma matrices ($\Gamma_M$) as well as the chirality matrix ($\tilde
\Gamma$) can be chosen to be real. The same holds true for the $AdS_2$
part (for example $\gamma_0 = i \sigma_2$, $\gamma_1 = \sigma_1$ and
$\tilde \gamma = \sigma_3$). For $S^2$ the gamma matrices
($\sigma_\alpha$) can be chosen to be real (e.g.  $\sigma_1$ and
$\sigma_3$) while the chirality matrix ($\tilde \sigma$) is imaginary
($\sigma_2$) and in 6 Euclidean dimensions all the gamma matrices ($\gamma_m$),
including the chirality matrix can be chosen purely imaginary. With
these conventions we construct ten-dimensional gamma matrices as
\begin{eqnarray}
  \label{g10}
  \Gamma_i & = &\gamma_i \otimes \mathbf{1} \otimes \mathbf{1} \; ,
  \nn \\
  \Gamma_a & = &\tilde \gamma \otimes \sigma_a \otimes
  \mathbf{1} \; , \\
  \Gamma_m & = & \tilde \gamma \otimes \tilde \sigma \otimes
  \gamma_m\; . \nn 
\end{eqnarray}
The definitions of the chirality matrices are
\begin{eqnarray}
  \label{g5}
  \tilde \gamma & = & \tfrac12 \epsilon^{i j} \gamma_i \gamma_j \;
  \qquad \tilde \sigma = \tfrac{i}2 \epsilon^{a b}
  \sigma_{a} \sigma_{b} \; , \\
  \tilde \gamma_7 & = & \tfrac{i}{6!} \epsilon^{m_1 \ldots m_6}
  \gamma_{m_1} \ldots \gamma_{m_6} \; , \qquad \tilde \Gamma =
  \tfrac{-1}{10 !} \epsilon^{M_1 \ldots M_{10}} \Gamma_{M_1} \ldots
  \gamma_{M_{10}} \; . \nn
\end{eqnarray}
With these definitions, we can check that
\begin{equation}
  \label{g11}
  \tilde\Gamma = \tilde \gamma \otimes \tilde \sigma \otimes \tilde
  \gamma_7 \; .
\end{equation}

\subsection{Some Gamma algebra}

Consider the 10 dimensional  metric of the form $ds^2= e^{2A}(g_{ij}dx^i dx^j+g_{ab}dx^adx^b)+h_{mn}dy^m dy^n$. 
In terms of the lower-dimensional gamma matrices, the antisymmetric product of 10d gamma matrices are given by
\begin{eqnarray}
\Gamma^{i m}&=&\gamma^i\wt\gamma\otimes\wt\sigma\otimes\gamma^{m}, \quad\quad \Gamma^{a m}=\wt\gamma^2\otimes\sigma^a\wt\sigma\otimes\gamma^{m},
\end{eqnarray}
\begin{eqnarray}
\Gamma^{ij p}&=&\gamma^{ij}\wt\gamma\otimes\wt\sigma\otimes\gamma^p, \quad\quad \Gamma^{ab p}=\wt\gamma\otimes\wt\sigma^{ab}\wt\sigma\otimes\gamma^p.
\end{eqnarray}

Some quantities used in our calculations involve the following:
\begin{eqnarray}
\left[\Gamma^{im}, \Gamma^{j n}\right]&=&\gamma^i \gamma^j \otimes \textbf{1} \otimes \gamma^m \gamma^n - \gamma^j \gamma^i \otimes \textbf{1} \otimes \gamma^n \gamma^m,\\
\left[\Gamma^{am}, \Gamma^{bn}\right]&=& \textbf{1} \otimes \sigma^a \sigma^b \otimes \gamma^m \gamma^n - \textbf{1} \otimes \sigma^b \sigma^a \otimes \gamma^n \gamma^m,\\
\left[\Gamma^{1m}, \Gamma_1\Gamma^{01n}\right]&=& \left[\Gamma^{0m}, \Gamma_0 \Gamma^{01n} \right] = 2
\gamma^{01} h^{mn} \textbf{1} \otimes \textbf{1} \otimes \textbf{1},\\
\left[\Gamma^{3m}, \Gamma_3 \Gamma^{23n} \right] &=& \left[\Gamma^{2m}, \Gamma_2 \Gamma^{23n} \right] = 2 \gamma^{mn} \textbf{1} \otimes \sigma^{23} \otimes \textbf{1},\\
\left[\Gamma_0 \Gamma^{01m}, \Gamma_1 \Gamma^{01n}\right] &=& 2 \gamma^{01} h^{mn} \textbf{1} \otimes \textbf{1} \otimes \textbf{1}, \\
\left[\Gamma_0 \Gamma^{23m}, \Gamma_1 \Gamma^{23n}\right] &=& -2 \gamma_{01} h^{mn} \textbf{1} \otimes \textbf{1} \otimes \textbf{1}, \\ 
\left[\Gamma_0 \Gamma^{01m}, \Gamma_1 \Gamma^{23n} \right] &=& -\left[\Gamma_0 \Gamma^{23m}, \Gamma_1 \Gamma^{01n} \right] = 2 \gamma^{mn} \textbf{1} \otimes \sigma^{23} \otimes \textbf{1}.
 \end{eqnarray}

\subsection{$SU(3)$ representations of torsion}
In a three-dimensional complex manifold with $SU(3)$ structure, there are two globally defined spinors constant under a connection $\nabla^{(T)}$ with torsion. This connection splits in terms of the Levi-Civita connection $\nabla$ and a contorsion component $\kappa$. The contorsion $\kappa$ is classified according to its $SU(3)$ representations.
We can elucidate these representations by computing the
covariant derivatives of the K\"ahler  $(1,1)$-form $J$ and the holomorphic $(3,0)$-form $\Omega$.  These forms can be written in terms of the internal spinor as $J_{mn}=\eta^\dag \gamma_{mn} \eta$ and $\Omega_{mnp}=\eta^\dag_+\gamma_{mnp}\eta_-$. For an internal manifold with $SU(3)$ holonomy, the internal spinor is covariantly constant and $\nabla_qJ_{mn}=\nabla_q\Omega_{mnp}=0$.\\

In general, for $SU(3)$ structure, $dJ$ is a threeform composed of all possible combinations of complex forms. Therefore, $dJ$ transforms as ${\bf 20}$ of $SU(3)$ which decomposes as
\begin{eqnarray}
{\bf 20}= ({\bf 1}\oplus {\bf 1})\oplus ({\bf 6}\oplus{\bf \bar{6}})\oplus ({\bf 3}\oplus{\bf \bar{3}}),
\end{eqnarray}
corresponding to $[(3,0), (0,3)]$, $[(2,1),(1,2)]$ and $[(1,0), (0,1)]$ forms, respectively. Such representations are labeled as ${\cal W}_1, {\cal W}_3$ and ${\cal W}_4$. The ${\cal W}_3$ representation corresponds to traceless forms.\\

Similarly, $d\Omega$ is a fourform in the ${\bf 24}$ of $SU(3)$. Since $d\Omega= (d\Omega)^{3,1}+(d\Omega)^{2,2}+(d\Omega)^{0,0}$, the ${\bf 24}$ representation decomposes, respectively, as
\begin{eqnarray}
{\bf 24}=({\bf 3}\oplus {\bf 3})'\oplus ({\bf 8}\oplus{\bf 8})\oplus ({\bf 1}\oplus{\bf 1}).
\end{eqnarray}
These representations are accordingly labeled by ${\cal W}_5, {\cal W}_2$ and ${\cal W}_1$. The ${\cal W}_2$ representation corresponds to traceless $(2,2)$-forms.

\bibliography{RB}
\addcontentsline{toc}{section}{Bibliography}
\bibliographystyle{TitleAndArxiv} 
\end{document}